\documentclass{aa}

\usepackage{graphics}
\usepackage{times}

\def\te{T_{\rm eff}}
\def\wig#1{\mathrel{\hbox{\hbox to 0pt{%
\lower.5ex\hbox{$\sim$}\hss}\raise.4ex\hbox{$#1$}}}}
\def\msol{\mbox{M}_\odot}
\def\mbol{\mbox{M}_{\rm bol}}
\def\mv{\mbox{M}_V}
\def\minf{m_{\rm inf}}
\def\te{t_e}
\def\kms{\,\mbox{km.s}^{-1}}
\def\mssurf{\mbox{M}_\odot\, \mbox{pc}^{-2}}
\def\msvol{\mbox{M}_\odot\, \mbox{pc}^{-3}}
\def\vrot{v_{\rm rot}}
\def\beq{\begin{equation}}
\def\eeq{\end{equation}}

\def\dx{\; {\mathrm d}x} \def\dm{\; {\mathrm d}m} \def\dv{\; {\mathrm
d}v} \def\d{\; {\mathrm d}} \def\rmP{{\mathrm P}} \def\rmd{{\mathrm d}}
\def\inf{{+\infty}}

\begin{document}

\def\aj{AJ}                  
\def\araa{ARA\&A}             
\def\apj{ApJ}                 
\def\apjl{ApJ}                
\def\apjs{ApJS}               
\def\ao{Appl.Optics}          
\def\apss{Ap\&SS}             
\def\aap{A\&A}                
\def\aanda{A\&A}                
\def\aapr{A\&A~Rev.}          
\def\aaps{A\&AS}             
\def\aandas{A\&AS}             
\def\azh{AZh}                 
\def\baas{BAAS}               
\def\jrasc{JRASC}             
\def\memras{MmRAS}            
\def\mnras{MNRAS}             
\def\pra{Phys.Rev.A}          
\def\prb{Phys.Rev.B}          
\def\prc{Phys.Rev.C}          
\def\prd{Phys.Rev.D}          
\def\prl{Phys.Rev.Lett}       
\def\pasp{PASP}               
\def\pasj{PASJ}               
\def\qjras{QJRAS}             
\def\skytel{S\&T}             
\def\solphys{Solar~Phys.}     
\def\sovast{Soviet~Ast.}      
\def\ssr{Space~Sci.Rev.}      
\def\zap{ZAp}                 

\thesaurus{}

\title{Towards a consistent model of the Galaxy: I. kinematic
properties, star counts and microlensing observations}

\author{{\sc D. M\'era$^{1,3}$, G. Chabrier$^1$ and R. Schaeffer$^2$}}

\institute{$^1$ C.R.A.L. (UMR CNRS 5574),
Ecole Normale Sup\'erieure, 69364 Lyon Cedex 07, France,\\
$^2$Service de Physique Th\'eorique,
CEA Saclay, 91191 Gif-sur-Yvette, France\\
$^3$Physics department, Whichita State University, 1845 Fairmount, Wichita KS 67260, USA}

\date{Received date ; accepted date}

\maketitle

\markboth{}{M\'era, Chabrier and Schaeffer: Consistent model of the
Galaxy I}

\begin{abstract}

We examine the most recent observational constraints arising from
i) small-scale and large-scale Galactic dynamical properties, ii)
star counts of population I and II stars at faint magnitude and iii)
microlensing experiments towards the Large Magellanic Cloud and the
Galactic centre. From these constraints, we determine the halo and disk
stellar mass functions and stellar content down to the bottom of the main
sequence, which yields the normalization of the halo/disk {\it total}
stellar population, and we infer the contributions of sub-stellar objects
to the mass budget of the various Galactic regions.

The consistent analysis of star counts and of the {\it overall}
microlensing observations in the Bulge are compatible with a small
contribution of brown dwarfs to the Galactic mass budget $\rho_{\rm
BD}/\rho_* \leq 0.2 $. However the {\it separate} bulge/disk analysis
based on the bulge clump giants is compatible with a substantial
population of disk brown dwarfs, $\Sigma_{\rm BD}/\Sigma_*\leq 1 $. More
statistics of microlensing events towards the Galactic center and a
better determination of the velocity dispersions in the bulge should
break this degeneracy of solutions.

For the halo, we show that a steep mass-function in the dark halo is
excluded and that low-mass stars and brown dwarfs represent a negligible
fraction of the halo dark matter, and thus of the observed events
towards the LMC. The nature of these events remains a puzzle and halo
white dwarfs remain the least unlikely candidates.


\end{abstract}

\section{Introduction}

There is compelling evidence for believing that most of the  matter in
the Universe is under the form of dark, yet unobserved components. The
observed density of baryons in galaxies $\Omega_{\rm star+gas} \sim
0.003 \, h^{-1}$ ($h$ is the Hubble constant in units of 100 $\kms \,
\mbox{Mpc}^{-1}$) may  represent only a fraction of the value predicted by
primordial nucleosynthesis, $\Omega_B\approx 0.01 \, h^{-2}$ (Copi, Schramm
\& Turner, 1995)\nocite{Copietal95}. There is also evidence
that spiral galaxies are surrounded by a large amount of non-luminous
mass of unknown nature. These two facts suggest that baryonic dark
matter is a possible candidate for halo dark matter. A breakthrough
in this longstanding, unsolved problem has been accomplished recently
with the developement of microlensing experiments, by inferring the
presence of dark objects through their gravitational effect on luminous
matter. The analysis of the EROS (Aubourg et al., 1993; Ansari et al.,
1996)\nocite{Aubourgetal93,Ansarietal96} and MACHO (Alcock et al., 1993;
1996)\nocite{Alcocketal93,Alcocketal96lmc} observations towards the LMC,
in particular, reveal the presence of {\it some} baryonic matter in
the Galactic halo, although the inferred optical depth shows that this
dark baryonic matter probably does not provides {\it all} the sought
missing mass. On the other hand, microlensing experiments towards the
central regions of the Galaxy yield a mass density of unseen star-like
objects about three times larger than the value expected from standard
disk+bulge models (Udalski et al. 1994;\nocite{UdalskiOgle94e} Alcock et
al., 1997).\nocite{Alcocketal97}  These two results from microlensing
experiments - the lack of events in the halo and the excess of events
in the bulge - lead to the tempting conclusion that more (resp. less)
galactic dark matter than expected previously resides in the disk
(resp. the halo). On the other hand, severe constraints on the amount
of dark matter in the disk and in the halo arise from the observed
large-scale kinematic properties, i.e. the rotation curve of the
Galaxy at distances larger than the observed luminous distribution,
as well as from the small-scale dynamical properties, i.e. the stellar
velocity dispersion in the solar neighborhood.  This latter information
is derived from the measurement of the vertical acceleration due to
the galactic potential near the Sun, which yields the determination
of the local dynamic surface density. Finally, star counts have now
been obtained at very faint magnitudes, either from the HST or from
ground-based deep-magnitude surveys, and provide stringent constraints
on different parameters entering galactic modeling such as scale lengths,
scale heights, and more indirectly stellar mass functions.

These two latter type of constraints, galactic dynamics and star counts,
have usually suggested that most of the Galactic missing mass resides
in the outer so-called dark halo. There is then an apparent conflict
between the afore-mentioned microlensing results and standard astronomical
observations. This paradigm stresses the need to reconsider the standard
model of the Galaxy, where all dark matter resides in the halo, and
to derive a Galactic mass-model consistent with the three types of
observational constraints, kinematics, star counts and microlensing. This
is the aim of the present study.

The calculations will be presented in two joint papers. In the present one
(Paper I), we examine in detail all observational constraints arising
from the most recent determinations of small-scale and large-scale
Galactic dynamic properties (\S 2), from star counts in the Galactic disk,
bulge and spheroid (\S 3), and from microlensing experiments towards both
the Large Magellanic Cloud (LMC) and the Galactic center (\S 4). A {\it
consistent analysis} of all these observations yields the determination
of the stellar mass functions, slope, normalization and minimum mass,
and thus of the amount of mass under the form of stellar and sub-stellar objects
in the various regions of the Galaxy, disk, bulge, spheroid and dark halo.

In paper II, these results will be used to derive a consistent model for
the Galaxy, confronted to all types of observational constraints.  We
will discuss in detail the two possibilities of an essentially non-baryonic
and a dominantly baryonic mass model for the Galaxy (M\'era, Chabrier \&
Schaeffer, 1997).\nocite{Meraetal97b}

\section {Kinematic constraints}

In this section we examine the most recent observational determinations of
the kinematic properties of the Galaxy. We first consider the asymptotic
circular velocity, in the Milky Way and in other spiral galaxies, and
then the velocity distribution in our Galaxy up to 60 kpc. Different
determinations of the total {\it mass} of the Galaxy, converted
into equivalent circular rotation velocities $v_{rot}^2=GM(r)/r$,
are also examined. In the second part of this section, we focus on the
determination of the local surface density, inferred from the measurement
of the gravitational acceleration of the Galactic potential near the
Galactic mid-plane. From all these observations, we derive the most
likely values for the Galactic mass, rotation velocity, and amount of
dark matter near the Sun, to be reproduced by the final galactic model.

\subsection{Large scale velocities}

\subsubsection{Circular velocity}

The most recent determinations of the circular velocity as a function
of the Galactic radius have been reviewed by Fich and Tremaine
(1991).\nocite{FichTremaine91} Their compilation of the measurements
of the outer rotation curve from CO, HI and HII observations yields
the accurate determination of a velocity of about 220 $\kms$, constant
within $\sim$10\% up to 14 kpc, with extremely small error bars (a few
\%). At larger distance there is a hint for a small increase of the
rotation curve, but the uncertainty in the data does not allow any
robust conclusion.
In any case,
there is no sign for a significant decrease of the velocity up to
about 20 kpc. More recent observations of neutral hydrogen out to 2.5
$R_\odot$, where $R_\odot\sim 8$ kpc is the galactocentric position
of the Sun, yield a slightly lower value $\vrot=200\,\pm 10\,\kms$
(Merrifield, 1992).\nocite{Merrifield92} On the other hand, Schechter
et al (1989)\nocite{Schechter89} report $\vrot = 248\,\pm 16\,\kms$
from the kinematics of carbon stars in the outer Galaxy. As noted by
Kuijken \& Tremaine (1994),\nocite{KuijkenTremaine94} a difference
between the value determined from stellar and HI tracers is expected,
because of the ellipticity of the galactic potential. This will decrease
the afore-mentioned stellar value.

The determination of the rotation curves of external galaxies (Casertano
and Van Albada 1990)\nocite{CasertanoAlbada90} characterizes what these
authors call ``bright galaxies'', with circular velocities bracketted
between 180 and 260 $\kms$. These rotation curves show occasionally a
slight drop beyond 20 kpc, which for all of them remains within less
than 20\% at 35 or 40 kpc, almost within the error bars. The rotation
curves are thus flat within the error bar determination ($\sim 10-20\%$)
up to the furthest distance at which hydrogen is detectable, i.e more
than twice the radius of the visible stellar component. Although no
rotation velocity measurements exist for our Galaxy beyond 19 kpc, Fich
and Tremaine (1991)\nocite{FichTremaine91} argue that, by comparison
with other galaxies, our rotation curve should extend similarly up to
at least 35 kpc.

Note that the Galactic rotation is well confined to the disk plane. At
1 kpc above the disk, metal-rich stars show similar rotation velocities
and are considered as belonging to the (thick) disk population. Most of
the metal-depleted stars ($[Fe/H] < -1.5$), which probe essentially the
halo population, rotate only with $\sim 30-40\, \kms$ assuming a
rotation velocity 220 $\kms$ for the Local Standard of Rest, arguing
for a non-rotating, or slowly rotating halo (Beers and Sommer-Larsen,
1995).\nocite{BeersSommer-Larsen95}

\subsubsection{Velocity distribution up to 60 kpc}

Attempts to determine the gravitational potential of our Galaxy at large
distances, from the study of the motion of globular clusters, require some
modelling of the velocity distribution in the Galactic halo, which enters
the Jeans equation (Binney \& Tremaine 1987).\nocite{BinneyTremaine87}
The compilation of Harris and Racine (1979)\nocite{HarrisRacine79}
led Frenk and White (1980)\nocite{FrenkWhite80} to use a 66 cluster
sample, with various assumptions for the shape of the Galactic halo,
to determine the mass of the Galaxy within 33 kpc. These clusters show
an overall $60\pm 26\,\kms$ circular rotation (assuming $v_{lsr}=220\,\kms$), again advocating for
a slow rotation of the halo, but also for important non-circular motion.
Translating the dynamically measured mass of the halo within a given
radius into an effective circular rotation velocity ($\vrot^2=GM/R $), their
results can be summarized as follows : this velocity always lies within
the range $200\le v \le 319\,\kms$, for {\it all} models, at the 90\%
confidence level. With the additional constraint that the globular cluster
distribution is not more flattened in the outer region than in the inner
region, the
allowed velocity-range reduces to 200-265 $\kms$. A second restriction
stems from the fact that if the velocity distribution is not isotropic,
the radial rms velocity $\sigma_r$ must be larger than the tangential velocity
$v_t$, because of infall towards the centre of the Galaxy. 
This leads to the final range $224 \kms < \vrot(33\, \mbox{kpc}) <
265 \kms$, the uncertainty being due to systematic errors.

The more distant the constraint, the more relevant it is to determine
the dark matter content of the Galaxy, but the more uncertain
the data.  The sample of 17 globular clusters and dwarf spheroidal
galaxies at galactocentric distances from 20 up to 60 kpc has been
analysed by Hartwick and Sargent (1978).\nocite{HartwickSargent78}
They determine the mass within 60 kpc, under the two opposite extreme
assumptions of purely radial or purely isotropic velocity distributions,
respectively. Translated into a circular velocity, this leads respectively
to $\vrot = 159 \pm 30 \kms$ and $\vrot= 237 \pm 30 \kms$. Caldwell
and Ostriker (1981)\nocite{CaldwellOstriker81} used the more realistic
geometrical mean of the two values $\vrot = 192 \pm 42 \kms$. The latter
finally has been scaled by Bahcall et al (1983)\nocite{Bahcalletal83}
to $\vrot(60\, \mbox{kpc}) = 205 \kms$ (to which must be added a $\pm 45\, \kms$
uncertainty) in order to take into account more recent determinations of
the local parameters near the Sun. These results are consistent with the
more recent and much more precise (although requiring a modelling of
the galaxy profiles) determination of the mass of the Galaxy by Kochanek
(1996),\nocite{Kochanek96} i.e.  $3.8\times 10^{11}\,\msol < M_{30}
< 5.4\times 10^{11}\,\msol$ and $4.0\times 10^{11}\,\msol < M_{60} <
8.7\times 10^{11}\,\msol$ at the 90\% C.L., where $M_{30}$ and $M_{60}$
denote the mass within 30 and 60 kpc, respectively.  This yields, within
the $1\sigma$ error level we have been using all along the present
analysis, $\vrot(30)=257\pm8\,\kms$ and $\vrot(60)=210\pm 13\,\kms$.

Fich and Tremaine (1991)\nocite{FichTremaine91} discuss the implications
of the velocities measured for the ``Magellanic Stream'', a hydrogen
bridge between the Magellanic Clouds and the Galaxy. A detailed
modelling shows that the $r^{-2}$ ($r^{-1.8}$ fits better) halo should
extend up to at least 60 kpc.  Although such a conclusion is reached
by means of several ``natural'' assumptions, and may be questioned,
the raw data show gas velocities around 200 $\kms$.

Peebles (1989, 1990, 1994)\nocite{Peebles89,Peebles90,Peebles94} has
examined the total mass of the Local Group, dominated by Andromeda and
the Milky Way. He finds total masses around $4\times 10^{12}\,\msol$,
suggesting a total mass of about $2\times 10^{12}\,\msol$ for our Galaxy.
Moreover, a modelling of all objects at a distance of 4 Mpc from the
Galaxy shows indeed that their velocity distribution is consistent with
the mass being attached to the objects.

\bigskip

In summary, all these studies lead to the conclusion that the Milky Way
has a mass $M_G\approx 2\times 10^{12}\,\msol$, with a nearly constant
rotation velocity $\vrot=220\,\pm 20\,\kms$, which implies an extension
of a Galactic dark halo up to at least 100 kpc.  This undetected mass
has to be compared with the mass under the form of visible matter, $\sim
10^{11}\,\msol$, which sets the scale for the amount of dark matter in
the Galaxy.

\subsection{Local dynamical surface density}

An important constraint on the mass density of the Galactic disk in
the solar neighbourhood is obtained from the study of the vertical
acceleration $K_z$ by means of the local stellar velocity distribution.
$K_z(z)$ is related to the (measured) vertical velocity dispersion
$\langle  v_z^2\rangle$ and density of one tracer population (Binney \&
Tremaine,1987)\nocite{BinneyTremaine87} by:

\beq K_z(z)=-\frac{1}{\rho}\frac{\partial \rho \langle  v_z^2\rangle}{\partial z} \eeq

By Taylor-expanding the Galactic potential above the disk, $\Phi(z)\approx
K_zz+Fz^2$, the Poisson-Jeans equation yields (Binney \& Tremaine
1987):

\beq K_z(z)={\partial \Phi \over \partial z} = K_{z0} + 2z F \label{Kz} \eeq

\noindent where the coefficient $K_{z0}$ depends on the mass content in
the disk (assumed to be infinitely thin, which means the above expression
is good for $ z \wig > 300$ pc only) while $F=2\pi G \rho_{halo}$ is the
halo contribution. The local surface density $\Sigma_\odot$ in the disk
is related to $K_{z0}$ by:

\beq 
\Sigma_{\odot} = \frac{K_{z0}}{2\pi G}  
\eeq

The surface density near the Galactic mid-plane determines
the mass of the disk, for a given disk scale length $R_d$:

\beq M_{\rm disk}\approx 2\pi \Sigma_{\odot} R_d^2\, e^{R_\odot/R_d}
\label{Mdisk} \eeq

In practice, $K_z(z)$ is measured at some height $z$ and must be corrected
from the halo contribution $2zF$ to determine $K_{z0}$ and $\Sigma_{\odot
\rm disk}$. Bahcall (1984a, 1984b),\nocite{Bahcall84a,Bahcall84b} in
agreement with Oort (1960)\nocite{Oort60} found $\Sigma_\odot \approx
70\,\mssurf$ with a claimed small error. Bienaym\'e, Robin and Cr\'ez\'e
(1987),\nocite{BienaymeRobinCreze87} using a sampling extending to higher
latitudes, obtained a better determination of $K_z(z)$ from 100 pc to
1 kpc, based on a modelling of the disk and the halo stellar population
arising from available observations. They found a lower surface density
for the disk,

\beq\Sigma_\odot = 50\, \mssurf ,\eeq

\noindent with a $10\,\mssurf$ uncertainty, and stressed that the
uncertainties in Bahcall's extrapolation to low $z$ were underestimated.
These authors, however, actually measure the acceleration at 1 kpc, where
the surface density is essentially the {\it total} surface density, and
get $K_z(1\,{\rm kpc})/2\pi G\approx 70 \,\mssurf$. The afore-value of
$\Sigma_{\odot}$ quoted for the disk is obtained after substracting
the standard halo contribution\footnote{Throughout this paper,
the {\it standard} halo is defined as a halo with a density profile
$\rho(r)=\frac{\vrot^2}{4\pi G} \frac{R_\odot^2 + R_c^2}{r^2+R_c^2}$,
with $\vrot=220\,\kms$ and $R_c=5$ kpc, see paper II. Note that this
halo is slightly heavier than the one used for instance by the MACHO
group, which corresponds to $\vrot = 204\, \kms$.} (see eqn.(2)), and
thus becomes model-dependent. Subsequently, Kuijken and Gilmore (1989,
1991),\nocite{KuijkenGilmore89a,KuijkenGilmore89b,KuijkenGilmore89c,KuijkenGilmore91}
found an observed {\it total} density $\Sigma(1.1\,\rm kpc)=71 \pm 6$
$\mssurf$ and argued that, after a theoretical correction for the
contribution of a standard massive halo similar to the one used by
Bienaym\'e et al. (1987),\nocite{BienaymeRobinCreze87} the surface
density of the disk should be reduced to :

\beq\Sigma_\odot=48\pm 9 \mssurf,\eeq

\noindent in remarkable agreement with the afore-mentioned determination.
A subsequent reanalysis of the Kuijken \& Gilmore (1989) work gives in fact $\Sigma_\odot=54\pm 8\,\mssurf$ (Gould, 1990).
We stress, however, that the two afore-mentioned surface densities at 300
pc depend somehow on the assumption made for the dark halo contribution
(Eq. (\ref{Kz})).

More recently, Bahcall, Flynn and Gould
(1992),\nocite{BahcallFlynnGould92} using the new data of Flynn
and Freeman (1993),\nocite{FlynnFreeman93} re-analyzed Bahcall's
(1984)\nocite{Bahcall84a,Bahcall84b} previous determination, and obtained

\beq\Sigma_\odot = 85 \pm 25 \mssurf.\eeq

The new lower limit weakens substantially the conflict with
the afore-mentioned values, at the price however of a large
uncertainty. The most recent analysis of these data by Flynn and Fuchs
(1994),\nocite{FlynnFuchs94} who added a new normalization point,
yields a best fit model:

\beq\Sigma_\odot = 52 \pm 13\,\,\mssurf\label{SigFF}\eeq

\noindent Both groups include a standard halo contribution in their
determination but for the average height of their sample ($\sim
300$ pc), the halo correction is small ($\sim 6\,\mssurf$) and
the associated uncertainty lies well within the observational error
bars. The reason why Bahcall et al. (1992)\nocite{BahcallFlynnGould92}
considered their value to disagree significantly with the Bienaym\'e
et al. (1987)\nocite{BienaymeRobinCreze87} and Kuijken \& Gilmore
(1991)\nocite{KuijkenGilmore91} values is that even with a $1\sigma$
deviation, the probability for $\Sigma_\odot$, as given by (7), to be
less than $60\, \mssurf$ is only 7\% and thus has only one chance in 14 to
occur, as stated by these authors. More optimistically, comforted by the
recent Flynn \& Fuchs (1994)\nocite{FlynnFuchs94} result, we note that
all these determinations of $\Sigma_\odot$ should rather be considered
as consistent at the $1\sigma$ level: the uncertainty-weighted average
of these four different results (5)-(8) yields:

\beq\Sigma_\odot = 51 \pm 6\,\,\mssurf,\label{SigA}\eeq 

\noindent very close to Eq. (\ref{SigFF})\footnote{Strickly speaking, the Bahcall et al. (1992) and the Flynn \& Fuchs (1994) values are not completely independent but the weight of the Bahcall et al. value is small. In any event, the value used as a reference all along our calculations is the most recent Flynn \& Fuchs one.}. We thus may consider the value
(\ref{SigFF}) as the most accurate present determination of the solar
dynamical density.

\section{Star counts}

\subsection{Disk}

The determination of the faint end of the stellar luminosity function
(LF), an essential issue to infer the low-mass star and the sub-stellar
contributions to the Galactic mass budget, has also been subject to strong
debate. Wielen et al. (1983)\nocite{Wielenetal83} found that, whereas the
LF was steadily rising up to $\mv =13$, there was a dip beyond this value,
leaving very little possibility for a substantial contribution of low-mass
stars to the mass of the disk. Subsequent studies (Stobie et al. 1989,
Tinney et al. 1992)\nocite{StobieIshidaPeacock89,TinneyMouldReid92}
showed that the dip is even more pronounced and starts already at $\mv =
12$. Gould, Bahcall and Flynn (1996;1997),\nocite{GouldBahcallFlynn97} using
HST observations at faint magnitudes, confirm a dropping LF beyond
$\mv=12$, which corresponds to $m\sim 0.25\,\msol$, although the last
bin of the HST LF clearly shows an upturn at $\mv > 16$.
Most of these LFs are based
on a photometric determination of the distance, which thus relies
on a color-absolute magnitude relationship, up to about a few hundred
pc. On the other hand, Kroupa (1995)\nocite{Kroupa95} used a nearby LF,
allowing a {\it geometric} determination of the distance. This yields
larger statistical errors, due to the number-limited sample, but better
systematic corrections, in particular the ones arising from the unresolved
binary companions. The nearby LF agrees with the photometric one up to $\mv \sim 12$ but is essentially flat beyond this limit.
Doing a careful correction for the Malmquist bias and
for the presence of unresolved binaries, Kroupa, Tout and Gilmore (1990; 1993)\nocite{KroupaToutGilmore90,KroupaToutGilmore93}
and Kroupa (1995)\nocite{Kroupa95} have shown that the properly
corrected ground-based photometric LFs are consistent with the local
geometric LF.

The issue, however, remains unsettled for the HST LF (Gould et al.,
1996; 1997). The HST is almost not subject to Malmquist bias since its
limit magnitude allows the observation of the bottom of the main sequence
up to the edge of the disk, but it is subject to uncompleteness due to
unresolved systems near the faint end\footnote{The HST is insensitive to
binaries with separations $\wig < 0.^{''}3$ and thus misses essentially
all secondaries in late M-dwarf binary systems.}. Gould et al. (1997),
however, found that even with this correction, their LF is inconsistent
with the nearby LF. Near the end of the LF, the binary correction is,
according to these authors, at most a factor 2, whereas a factor $\sim 5$
is required to reconcile the two kinds of LFs. Therefore, unresolved binaries are certainly not the source of the discrepancy between the two LF's. An important quantity to calculate correctly the correction due to binaries
is the rate of binaries and the mass-distribution of the secondaries.
An accurate determination of such
quantities require decade-long observations, as conducted for example by
Duquennoy and Mayor (1991; DM91) for nearby F and G stars, with a fairly
accurate sensitivity ($\sim 300\, \mbox{m.s}^{-1}$). Note that, altough
there are severe natural observational biases toward the detection of
near equal-mass systems either from imaging or from spectroscopic surveys,
the completeness and the uniformity of the DM91 sample render it almost
without any observational selection effect, yielding an almost unbiased
mass-ratio determination for F and G star binaries.  The main conclusion
of the DM91 study is that a fairly large number ($\sim 60-70\%$) of single
stars form a multiple system, and that the shape of the mass-distribution
increases toward small mass ratios ($q=m_2/m_1<1$; see their figure
10). \nocite{DM91}

M-dwarf surveys to address the same issue are still in their infancy and
suffer from severe incompleteness and observational biases, so that it
will require more years to do a similar analysis for M-dwarf binaries.
In any case, as mentioned above, even a full knowledge of the binary correction will not account
for all the discrepancy between photometric and geometric LFs. For
this reason, we have considered in the present paper both (nearby
and HST) LFs.

\begin{figure} \resizebox{\hsize}{!}{\includegraphics{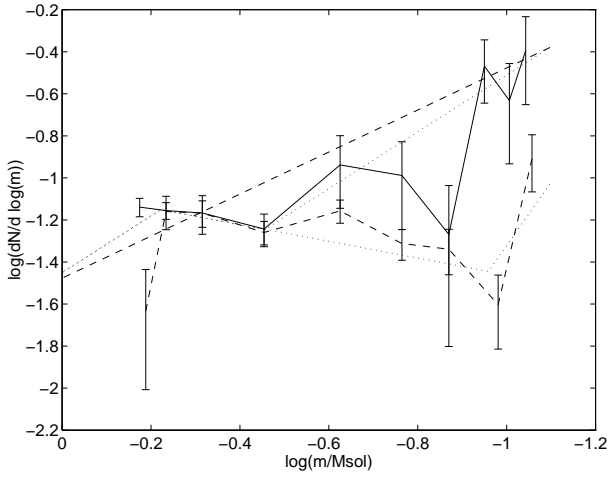}}
\caption[]{Mass functions derived from Kroupa (1995; solid line) and
Gould et al. (1997; dashed line) LFs. Also shown are the fits for both
MFs, which are identical for $m>0.35\msol$ (dotted lines). See table
\protect\ref{tablemf} for the parameters of the fits. The dashed line
is an overall fit given by Eq. (\ref{DiskMF}) with $\alpha =2$, which reproduces
reasonably well the MF derived from Kroupa's LF (note that there are only 2 stars in the lowest bin of Kroupa's LF; see text).\label{figmf}}
\end{figure}

We have converted both LFs into {\it mass-functions}
(MF), using low-mass star models which accurately reproduce observed
mass-M$_V$ relationships (Chabrier, Baraffe \& Plez, 1996). Both MFs are
displayed in Figure \ref{figmf} in a log-log scale.  The afore-mentioned
discrepancy between the two LFs is obvious on the MF, in particular
in the domain $\sim 0.1-0.25\,\msol$, with a factor $\sim 5$ ratio at
$\sim 0.1\,\msol$. 

Noting that the two lowest bins in the nearby LFs are the least
statistically significant (2 stars, see Kroupa (1995)), M\'era
et al. (1996a) parametrized the disk low-mass star MF in the solar
neighborhood down to the vicinity of the hydrogen-burning limit by the
following form:

\beq \mu(m) \approx 1.5 \pm 0.4 ({m \over 0.1\,\msol} )^{-2 \pm
0.5}\,\msol ^{-1}\, \mbox{pc}^{-3},\label{DiskMF}\eeq

The undetermination of the exponent reflects the fact that this fit is a
{\it reasonnable} overall power-law representation of the true MF but is
not perfect. As shown by Kroupa et al. (1993), the MF determined from the
nearby LF is better described by a series of segment power-law fonctions
$\mu(m)=dN/dm \propto m^{-\alpha}$.  We then have fitted the MFs derived
from the nearby and the HST LFs with such segmented power-laws. The
characteristics of these MFs are given in table \ref{tablemf} and
illustrated in Figure 1.  We will refer to these fits as MF (a) for Kroupa
(1995) and MF (b) for Gould et al. (1997).

\begin{table}
\caption{Fit of the two MFs considered in this paper. The MF (a) has
been derived from Kroupa (1995) LF, whereas the MF (b) has been derived
from Gould et al. (1997) LF. The MFs and their fits are displayed in
figure \protect\ref{figmf}. Each MF is considered to be a 3-segment
power-law $\mu (m)=dN/dm=A\, m^{-\alpha}$. Both MFs are
identical for $m>0.35$ M$_\odot$.  The mass is in $\msol$, the mass
function in M$_\odot^{-1}$ pc$^{-3}$. The upper mass limit has been set
arbitrarily to 10 $\msol$, since it has almost no influence on the mass
density. Note that 0.4 should be added to the slope derived from the HST LF for $m<0.6\,\msol$ to account for missed binaries (Gould et al., 1997)\label{tablemf}}
\begin{tabular}{cccc}
\hline
MF (a): & & & \\
mass & $\minf-0.35$ & $0.35-0.6$ & $0.6-10$\\
$\alpha$ & $2.35$ & $0.6$ & $2.35$ \\
$A$ & $5.98\times 10^{-3}$ & $3.76\times 10^{-2}$ & $1.54\times 10^{-2}$ \\
\hline
MF (b): & & & \\
mass & $\minf-0.11$ & $0.11-0.6$ & $0.6-10$\\
$\alpha$ & $4$ & $0.6$ & $2.35$ \\
$A$ & $2.08\times 10^{-5}$ & $3.76\times 10^{-2}$ & $1.54\times 10^{-2}$ \\
\hline
\end{tabular}
\end{table}

If we exclude the last bin, the HST MF corresponds to a slope $\alpha
< 1.0$ for $m/\msol \wig < 0.6$.
However, an extrapolation of the MF (ignoring the last bin) all the
way below 0.2 $\msol$ would correspond to about 2 stars in the last
bin whereas 10 are seen. The probability to see 10 stars whereas 2 are
expected is less than $5 \times 10^{-5}$. This suggests that the last
bin in the HST LF
is indeed significant, adding credibility for a rising MF near the
BD limit. As shown in Figure \ref{figmf}, the fit does not reproduce
exactly the last bin.  This takes into account a possible bias which
tends to overestimate the last bin and to underestimate the penultimate one
(Gould et al. 1997).

Indeed, for both MF's (nearby and HST), there is a hint for a rise near
0.1 Msol, suggesting the possibility for a substantial amount of brown
dwarfs in the disk.  We will take this as a basis for extrapolating
the stellar MF into the brown dwarf domain.  This is consistent with a
recent study by Mazeh, Latham \& Stefanik (1996) on the distribution of
secondary masses near the substellar limit.  Although this study relies on
the detection of 3 very-low-mass companions out of 20 F and G dwarfs and
subgiants, and thus preclude the precise determination of a substellar
MF, these detections suggest that the mass distribution (the primaries
of the sample all have $\sim 1\,\msol$) rises near the sub-stellar limit.

In any case, whereas the precise determination of the shape of the MF
near the bottom of the main sequence, which requires larger statistics,
is of prime importance for star-formation theory, the qualitative
behaviour, and its overall characterization, is what is most essential
to estimate the stellar and sub-stellar mass contributions to the disk
mass budget. Both rising MF's near the BD limit suggest a substantial
amount of substellar objects and thus a contribution to the disk mass
density. The extension into the brown dwarf domain will be analyzed in
the next section, in connection with microlensing observations.

The integration of the nearby MF displayed in Fig. \ref{figmf} gives
the local low mass star ($0.07 \msol \le m \le 0.65 \msol$) densities
$\rho_{\sc LMS} = 2.56\pm 0.25\times 10^{-2}\ \msvol$ in case (a),
and $\rho_{\sc LMS} = 1.43\pm 0.09\times 10^{-2}\ \msvol$ in case
(b). Since the HST LF does not include companions of multiple systems,
we have tentatively corrected for this bias on the basis of the Duquennoy
and Mayor (1991) observed mass ratio distribution mentionned above. The
correction is approximated by a a linear fonction, whose value is 10\%
at 0.6 $\msol$, and 80\% at 0.1 $\msol$. We renormalize the HST MF
to keep the same normalization at 0.6 $\msol$ with this correction.
The resulting density is $\rho_{\sc LMS} = 1.88\times 10^{-2}\ \msvol$,
almost in agreement with the nearby value. An average value is then:

$$ \rho_{\sc LMS} = 2.2\pm 0.3\times 10^{-2}\ \msvol $$

With a vertical sech$^2$ density profile of scale height $h_d\approx
320$ pc (Gould et al., 1997), the contribution of low-mass stars (LMS)
down to the bottom of the main sequence ($0.1\wig < m/\msol\wig < 0.6$)
to the disk surface density is then:

\beq \Sigma_{\rm lms}
\approx 14\, \pm \, 2 \, \mssurf\label{Siglms}\eeq

\noindent  Adding the contribution of more massive stars, obtained
with the MF and the luminosity-dependent scale height
determined by Miller \& Scalo (1979),\nocite{MillerScalo79} converted
into a mass-dependent scale height, yields the total contribution of
main sequence stars:

\beq \Sigma_\star =24\,\pm\, 3\, \mssurf,\label{Sigmastar}\eeq

Including the stellar remnants, white dwarfs (Liebert, Dahn and Monet,
1988) and neutron stars, $\Sigma_{\rm wd+ns} =$ 2 to 4 $\mssurf$,
and an estimated interstellar gas contribution $\Sigma_{\rm gas} =$
10 to 15 $\mssurf$ (Bahcall, 1984), the total surface mass density under
the form of directly detected baryonic components in the disk is thus:

\beq(\Sigma_\odot)_{\rm vis}=40\,\pm\,4 \,\mssurf\label{Sigobs1}\eeq

\noindent in good agreement with Gould et al. (1997).  The limits are
derived from very conservative estimates for the errors. 
Using a $\mbox{sech}
(|z|/z_0)^2$ vertical density distribution instead of an exponential
profile leaves these values almost unchanged.

\subsection{Bulge}

The mass and the shape of the Galactic bulge will be discussed in more
details in the next section. This mass is estimated to be $\sim
2\times 10^{10}\,\msol$, from dynamical considerations (Zhao et
al., 1996),\nocite{ZhaoSpergelRich96} with a size of about 1 kpc
(Kent 1992).\nocite{Kent92} The faint end of the LF in the Baade's
Window, down to $\mv \sim 10$, i.e. $m\approx 0.5\,\msol$, has
been studied with the HST Wide Field Camera (WFC) by Holtzman et al
(1993).\nocite{Holtzmanetal93} These authors find a Salpeter slope for the
MF $x \approx 2.35 \pm 1$, over the afore-mentioned mass range. There
is no indication for this MF to be different, except for its local
normalization, from the one determined for the disk, Eq. (\ref{DiskMF}).

\subsection{Thick disk}

A disk+bulge model does not reproduce successfully the observed
star counts and kinematic properties. This led Gilmore \& Reid
(1983)\nocite{GilmoreReid83} to introduce the so-called {\it thick disk},
with a scale height $h_{\rm td}\approx 1-1.5$ kpc, which seems to be
necessary for an accurate galactic modelling. Since the stellar velocity
dispersion $\langle v_z^2\rangle$ and the oblateness of the disk are
related, a {\it thick} disk, by definition, is less elongated than the
young (thin) disk, and has a larger velocity dispersion. This yields a
better agreement with observed velocity dispersions, and with star counts
in general. As shown recently by Reid et al. (1996)\nocite{Reidetal96}
and Chabrier \& M\'era (1997),\nocite{ChabrierMera97} a thick disk with
$h_{\rm td}\approx 1$ kpc and $q\approx 0.6$ is consistent with the
Hubble Deep Field counts at faint magnitudes. A larger scale height
and a spherically symmetric distribution predict substantially larger
star counts.  In spite of the small volume density of the thick disk
in the solar neighborhood, about a few percents of the young disk at 1
kpc (Gilmore, Wyse \& Kuijken; 1989),\nocite{GilmoreWyseKuijken89} its
contribution to the dynamics is not negligible since its surface density
$\displaystyle \Sigma_{\rm td}(R)=\int_{-\infty}^{+\infty} \rho_{\rm
td}(R,z)dz$ represents $\sim$ 10-20\% of the young disk contribution.

Taking into account the thick disk contribution, and using the parameters
derived by Gould et al. (1997), the disk stellar surface density is
raised by $\sim 3\,\mssurf$ to

\beq \Sigma_\star = 27\,\pm\, 4\ \mssurf,\label{Sigmastar2}\eeq

\noindent so that the detected surface density
(\ref{Sigobs1}) reads:

\beq(\Sigma_\odot)_{\rm vis}=43\,\pm 5\mssurf \label{Sigobs}\eeq

\subsection{Spheroid}

The spheroid is also called in the literature the {\it stellar}
halo. It is defined by a spherical density distribution decreasing
as $\rho(r)\propto r^{-3}$ (Hubble profile) or equivalently a
$r^{1/4}$, de Vaucouleurs profile. It differs from the central
bulge essentially by a lower metallicity, $[M/H]\approx -1.5$
(Monet et al., 1992; Leggett, 1992; Baraffe et al., 1995,
1997)\nocite{Monetetal92,Leggett92,Baraffeetal95,Baraffeetal97}
and larger velocity dispersions, $\sigma_R\approx 160$ $\kms$,
$\sigma_z=\sigma_{\theta}\approx 90\ \kms$ (Dahn et al. 1995; Beers \&
Sommer-Larsen, 1995).\nocite{Dahnetal95,BeersSommer-Larsen95} The mass
of the spheroid is $\sim 10^9 \,\msol$ (Bahcall, 1986)\nocite{Bahcall86}
so its contribution to the galactic {\it mass} is not important. Its
main contribution concerns the halo {\it star counts}.

Richer and Fahlman (1992)\nocite{RicherFahlman92} obtained for the
spheroid a steeply rising LF down to the end of the observations whereas
Dahn et al. (1995),\nocite{Dahnetal95} from observations of high-velocity
($v_{\rm tan}> 200\,\kms$) subdwarfs in the solar neighorhood, get a LF
which decreases at the faint end ($\mv \wig > 12$).

M\'era, Chabrier \& Schaeffer (1996)\nocite{MeraChabrierSchaeffer96}
and Chabrier \& M\'era (1997)\nocite{ChabrierMera97} (see also Graff \&
Freese, 1996)\nocite{GraffFreese96} have determined the MF of the spheroid
from these observed LFs and from mass-magnitude relationships derived from
LMS models which reproduce accurately the main sequences of metal-poor
globular clusters, observed with the HST deep photometry surveys,
down to the vicinity of the hydrogen burning limit (Baraffe et al.,
1997).\nocite{Baraffeetal97}  The spheroid MF in the solar neighborhood
deduced from the Dahn et al. (1995) subdwarf LF is reasonably well
parametrized as (Chabrier \& M\'era, 1997):

\beq\mu(m)=4.0 \pm 1.0\, 10^{-3}\, \left({m\over 0.1\,\msol}\right)^{-1.7
\pm 0.2}\, \msol^{-1}\, \mbox{pc}^{-3},\label{SpheMF}\eeq

Integration of this MF gives the halo main sequence ($m\le 0.8 \,\msol$
for $t\ge 10$ Gyr) stellar density in the solar neighborhood :

\beq\rho_{h_{\star}}\approx 1.0 \times 10^{-4}\,\msvol\label{rhosph}\eeq

As shown by Chabrier \& M\'era (1997),\nocite{ChabrierMera97}
star count predictions based on this MF are in excellent agreement
with the Hubble Deep Field observations at very faint magnitudes
both in the $V$ and $I$-bands. Comparison of this value with the
disk main-sequence stellar density determined in \S3.1 gives the
normalization of the spheroid/disk stellar population at 0.1 $\msol$,
namely $\rho_{\star_{sph}}/\rho_{\star_d}\sim 1/400$ (Chabrier \& M\'era,
1997).\nocite{ChabrierMera97}

\subsection{Dark Halo}

Throughout this paper, the term "dark halo" defines what is characterized
by the $\rho(r)\propto r^{-2}$ density-profile, a consequence of the
flat rotation curve in the outer part of the Galaxy (see \S2). The
existence of a halo in the Galactic structure is rendered necessary
i) to account for the total mass of the Galaxy (see \S2.1.2), since
the disk(s), bulge and spheroid total mass amounts at most to $\sim
10^{11}\,\msol$ and ii) for stability conditions (Ostriker and Peebles,
1973),\nocite{OstrikerPeebles73} although the presence of a central
bar weakens this latter argument. There is observational evidence that
galactic dark halos present some oblateness with $q\sim 0.6$
(Sackett et al. 1994)\nocite{Sackettetal94}.  Were the mass-distribution
of the dark halo 2-dimensional (i.e. under the form of a disk),
the surface density in the solar neighborhood, $\Sigma_\odot\approx
\vrot/2\pi G R_\odot = 210\,\mssurf$ would be a factor 3 larger than
the dynamical upper limit (see \S2.2) (see M\'era et al., 1997, paper
II). This condition, and the motion of galaxies and gas in the Local Group
(\S2.1) are strong arguments in favor of a large, 3-dimensional halo
around the centre of the Galaxy.

The MF in the halo is presently unknown. The observation of the bottom
of the stellar main sequence, $m\sim 0.1\,\msol$, at a distance of 20
kpc, about the limit of the spheroid, requires surveys down to apparent
magnitudes $I \wig > 28$ (Baraffe et al. 1997).\nocite{Baraffeetal97}
It requires also an excellent angular resolution ($< 0.1$ arcsec)
to distinguish stars from galaxies. This is the reason why early
attempts using ground-based observations were never brought to a
conclusive end. The recent Hubble Deep Field star counts (Flynn
et al., 1996; Reid et al., 1996; Mendez et al., 1996)
\nocite{FlynnGouldBahcall96,Reidetal96,Mendezetal96}
up to $I=28$ are insufficient to determine a halo LF, because of
the limited field of view, but represent a severe constraint on the
Galactic model. These faint magnitude star counts have been shown to be
entirely consistent with a thick disk + flattened spheroid population,
characterized by their respective I-(V-I) relationships and mass
functions (\ref{DiskMF}) and (\ref{SpheMF}) (Chabrier \& M\'era,
1997).\nocite{ChabrierMera97} As shown by these authors,
a (even flattened) dark halo with the afore-mentioned $1/r^2$
profile and MF (16) would predict at least 500 more stars than observed in the HDF
field of view. This shows that, if the dark halo LF is the same as the
spheroid one, there is a negligible stellar
population in the {\it dark halo}, less than 0.1\% of the dynamical mass.

A promising technique to determine more accurately the halo MF comes
from the recent surveys of the outer parts of nearby galaxies (Sackett
et al. 1994; Lequeux et al.  1996)\nocite{Sackettetal94,Lequeuxetal96}
and from the relation between the slope of the MF and the inferred colors
of the halo diffuse emission (M\'era, 1997).\nocite{Mera97}

Another important indication is given by the LFs of globular clusters. The
MFs of the different clusters observed with the HST have been determined
recently by Chabrier \& M\'era (1997).\nocite{ChabrierMera97} The typical
LFs, determined near the respective half-mass radii, are rising up to
$\mbol \sim 10$ and drop below. This is consistent with monotonically
{\it rising} MFs down to $\sim 0.1\,\msol$, with a very weak dependence on
metallicity, well described by power-law functions with $\alpha \sim 0.5-1.5$,
similar to the ones determined in the previous sections for the disk
and the spheroid.  Although mass segregation effects are expected to be
small near the half-mass radius (see e.g. King et al., 1995; Chabrier \&
M\'era, 1997),\nocite{King95,ChabrierMera97} the evolution of
the clusters, and their evaporation, might lead to {\it initial} MFs
(IMF) slightly steeper that the afore-mentioned determination. This
would yield even better agreement with the MF (\ref{SpheMF}) and would
be consistent with the suggestion that the formation of spheroid
field stars and globular clusters has occured in a similar manner and
time frame (Fall and Rees, 1985).\nocite{FallRees85}

Since the dark-halo stellar population is likely to be insignificant, as
mentioned above, Eq. (\ref{SpheMF}) gives the total halo (spheroid + dark
halo) stellar contribution to the dynamically-determined local density
$\rho_{\rm dyn}\sim 0.01\,\msol.pc^{-3}$, namely $\rho_{h_{\star}}/\rho_{\rm
dyn}\sim 1\%$. This puts more stringent limits than the ones obtained from
the HST counts alone (Flynn et al., 1996)\nocite{FlynnGouldBahcall96} and
shows convincingly that low-mass stars represent a completely negligible
fraction of the halo mass.

\bigskip

In summary, the star count analysis of different
stellar populations corresponding to different regions
of the Galaxy, based on accurate evolutionary models for
low-mass stars (Baraffe et al., 1997; Chabrier et al., 1996,
1997),\nocite{Baraffeetal97,ChabrierBaraffePlez96,ChabrierMeraBaraffe97}
leads to the determination, slope and normalization, of the MF in the
disk, the bulge and the spheroid, down to the bottom of the main sequence.
Surprinsingly enough, given the different metallicities in these regions
and the wide range of stellar masses considered, these MF's are very
similar and reasonably well described by power-law MFs $dN/dm\propto
m^{-\alpha}$ with $\alpha \approx 1.5-2$.  A flatter MF below $\sim
0.5\,\msol$ for the disk, as sugested by the HST LF, cannot be excluded
but a more precise determination requires better statistics at the
faint end of the LF.  These MFs, consistent with all observed LFs and
deep-photometry counts, yield a fairly well determined contribution of
{\it all main sequence stars} to the disk and halo {\it mass budgets},
the dark halo stellar contribution to the Galactic mass being essentially
insignificant. The determination of the amount of mass under the form
of {\it sub-stellar} objects requires the analysis of the microlensing
experiments.

\section{Microlensing}

Within the past few years the microlensing technique has been applied
succesfully to the search for dark matter, by inferring the presence of
dark objects in the Galaxy through their gravitational effect on luminous
matter, as proposed initially by Paczy\'nski (1986).\nocite{Paczynski86}
Several groups worldwide have carried out microlensing experiments to
determine the amount of dark matter either in the halo of the Galaxy
(EROS and MACHO experiments) or in its central parts (OGLE and MACHO
experiments). We examine below these various experiments and connect the
results with the afore-mentioned stellar MFs to determine the contribution
of {\it sub-stellar objects} to the galactic mass budget.

\subsection{Disk and Bulge}

Numerous events have been detected in the direction of the Galactic centre
by the OGLE (Paczy\'nski et al., 1996)\nocite{Paczynskietal96} and MACHO
(Alcock et al.  1997)\nocite{Alcocketal97} surveys. We focus here on
the 40 events\footnote{We have excluded the two events that fail MACHO cuts
(104-B and 111-B), two events suspected to be variable stars (113-C and
121-B) and the binary event which is not accounted for by the efficiency.}
obtained from the first-year analysis of the MACHO collaboration,
whose detection efficiency $\epsilon(t)$ is well determined (Alcock et
al., 1997).\nocite{Alcocketal97} The number of events observed by OGLE
during the first two years is too small and the attempt to use averages
would be dominated by small number statistics. At any rate, adding,
as an extreme value, the 9 OGLE events to the 40 MACHO ones would not
improve significantly the statistics.

The interpretation of the three large-time events ($t_e>75$ days) in
the MACHO survey is still unclear and will be discussed below. Note that
there is no event with $35<t_e<75$ days.

The number of observed events allows a statistical analysis from
their time distribution. The average effective time $\langle  t_e\rangle$
can be used to estimate the average lens mass $\langle  m\rangle$ and
thus the minimum mass down to which the mass function (\ref{DiskMF})
extends (see Appendix):

\beq \langle \sqrt m \rangle={c\over 2\sqrt{GL}} \langle t_e
\rangle \langle \frac{1}{v_\perp} \rangle^{-1} \langle \sqrt{x(1-x)}
\rangle^{-1}\label{sqrtm}\eeq where the averages are relative to the
effective probability distributions $P_{\rm eff}$ for each variable
(see Appendix). For a simple gaussian velocity distribution,
$\langle\frac{1}{v_\perp}\rangle^{-1} = \sigma \sqrt{\pi/2}$,
where $\sigma$ is the velocity dispersion. The geometric factor
$\langle\sqrt{x(1-x)}\rangle$ is approximatly 0.39 for a standard
(Bahcall \& Soneira, 1980 + Kent, 1992) disk+bulge model, and is not
very sensitive to the choice of the model. The mean duration of the 40
events, taking into account the experimental efficiency, is $\langle
t_e \rangle=16.8$ days (omiting the efficiency leads to $\langle t_e
\rangle=20$ days). Then a first estimate of the average lens mass is:

\beq \langle \sqrt{m}\rangle^2 = 0.15\,\msol \eeq \noindent which is
only a rough estimate, with a velocity dispersion $\sigma=100\kms$
for the bulge. Excluding the three longest events yields
$\langle\sqrt{m}\rangle^2 = 0.09\msol$. These estimates introduce the
dilemma that will be discussed in detail below: in the first case,
microlensing results are compatible with essentially no brown dwarf
in the disk/bulge, whereas in the second case, the mass function must
extend significantly in the brown dwarf domain.

As mentioned in the Appendix A.2, we stress that $P(\te)$ has a long power-law
tail at large times, since $\te \propto v_\perp^{-1}$, which yields
an unphysical divergence of the moments. Therefore, the statistics for
the mass of the microlensing events can {\it not} be evaluated correctly
from the calculation of the moments, as done sometimes in the literature,
but must be evaluated accurately by using the detailed time distribution
(Note, however, that the moments of $1/t_e$  behave correctly).

We have calculated the time distribution with a Monte-Carlo
algorithm (see Appendix), which determines the number of events:

\beq N_{th}=E\times \int_0^{+\infty} \epsilon (\te) {d\Gamma \over d\te}\,
d\te \label{Nth}\eeq

\noindent where $E$ is the total exposure (in star-years) and
$\epsilon$ the experimental efficiency factor (Alcock et al.,
1997).\nocite{Alcocketal97} Calculations have been done for a standard
disk model (Bahcall \& Soneira, 1980),\nocite{BahcallSoneira80} and a
Kent (1992)\nocite{Kent92} model for the bulge, for the two MFs
considered in \S 3.1. We have taken velocity dispersions $\sigma
\sim 100\,\kms$ for the bulge and $\sigma \sim 20\,\kms$ for the disk
($40\,\kms$ in the radial direction), respectively, with a linear
interpolation in the intermediate region located between $\sim 2$ and 3
kpc from the Galactic center. As will be shown below, the uncertainty in
the velocity distribution of the stars is the main cause of uncertainty
in the present calculations.

Our calculations take into account the motion of the Sun and of the
source star in the determination of the lens velocity as well as the
variation of the distance of the {\it source} stars in the disk and
the bulge (see Appendix). The distance of the source stars in the
bulge is typically 7.5 kpc from the Sun, as determined by Holtzman
et al. (1993)\nocite{Holtzmanetal93} from the observed LF in Baade's
Window. The time-distribution and number of events obtained from these
calculations, for a given model, are compared to the MACHO observations
(Alcock et al., 1997).\nocite{Alcocketal97} Since the efficiency does
not include binary events, we must exclude those events. However, only a
small fraction of the events are expected to be due to both components
of a binary system (1 out of 45 in the first year, a fraction which
seems to be confirmed by the following two years). In most cases, the
Einstein disks of the stars in a multiple system do not overlap, and can
in first approximation be treated independently. The time distribution is
therefore barely affected by the binarity. In any case the number of events has
to be corrected only slightly (typically a few percent). In this section,
we are mostly interested in the determination of the mass function in
the substellar regime, which is derived only from the time distribution
of the observed events. The consistency between models and observations
is analyzed with a Kolmogorov-Smirnov (KS) test\footnote{The KS test
does not give a confidence level, but rather an exclusion criterium. Its
significance function has a very rapid variation between 10\% (distributions
most probably different) and 90\% (distributions most probably identical),
and any value in between does not give a strong conclusion. But the KS
test is better than the $\chi^2$ test for low statistics, and does not
depend on any binning of the data.}.

\begin{figure} \resizebox{\hsize}{!}{\includegraphics{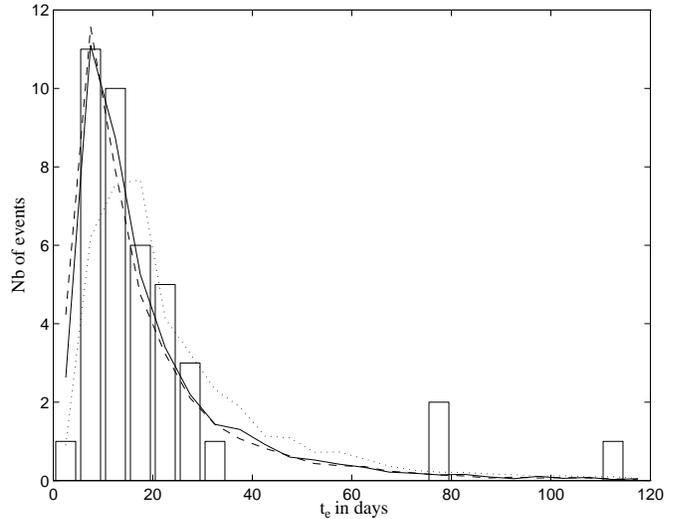}}
\caption[]{Observed time distribution (histograms) for all MACHO events
compared with the distribution of two models. Solid line: MF (a) with
$\minf=0.075\,\msol$, dashed line: MF (b) with $\minf=0.05\,\msol$,
and dotted line: MF (b) with $\minf=0.1\,\msol$. The latter case is
excluded by a KS test (see text), showing that brown dwarfs are required to explain
the observed distribution. \label{distte1}}
\end{figure}

We will first assume that the mass function in the bulge is the same
as in the solar neighborhood.  The mass function is extrapolated in the
brown dwarf domain down to a minimum mass, $\minf$, which is determined
when the best agreement between theory and observation is achieved. The
results are shown in Figure \ref{distte1}, for the mass functions (a)
and (b) defined in \S3.1, with $\minf = 0.075\, \msol$ and $\minf=0.05 \,
\msol$, respectively. The KS test gives a probability of 85\% for MF (a)
and only 45\% for MF (b). The value $\minf =0.05\, \msol$ for MF (b),
and its steeper slope in the brown dwarf domain renders MF (b) very similar to
a Dirac delta-function. Longer timescale events are then more difficult
to reproduce. On the other hand, for MF (a), the relative contribution
of higher masses is larger, with a better fit of the long timescale
observations.

We have also computed the time distribution in case (b) with
$\minf=0.1\msol$, which correspond to the absence of brown dwarfs
in the disk. This corresponds to the general decreasing behaviour
of the HST MF, ignoring the last bin. The time distribution, also
shown in Fig. \ref{distte1}, is incompatible with the observed time
distribution, with a KS result of 5\%, as already pointed out by Han
and Gould (1995). These results suggest that brown dwarfs are required
to explain all the 40 events of the MACHO first year bulge observations,
unless the effect of amplification bias (Alard, 1997; Han, 1997)\nocite{Han97} is important enough to account for the excess of short timescale events.

\bigskip

\begin{figure} \resizebox{\hsize}{!}{\includegraphics{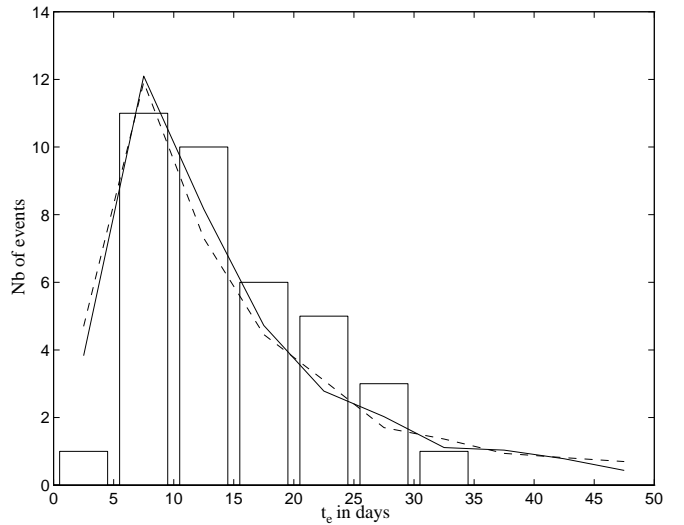}}
\caption[]{Same as figure \ref{distte1}, without the three longest
events (see text). Models
are: MF (a) with $\minf=0.056\,\msol$ (solid line), MF (b) with $\minf=0.047\,\msol$ (dashed line). \label{distte2}} 
\end{figure}

It has been suggested (Alcock et al 1997; Han and Gould,
1995)\nocite{Alcocketal97,HanGould95} that the large time events may
have another origin than main sequence stars, and might be due to
stellar remnants, white dwarfs or neutron stars. They may be due also
to rare statistical fluctuations due to the large tail of the time
distribution or, more speculatively, to dwarf nova eruptions (Della
Valle and Livio, 1996).\nocite{DellaValleLivio96}  If we exclude the 3
longest events from the present analysis, the agreement between theory
and observation improves significantly, as shown in Fig. \ref{distte2}
for the same MFs. This yields minimal masses for the two mass functions
$\minf=0.056\ \msol$ in case (a), and $\minf=0.047\ \msol$ in case
(b). The KS significance of the two MFs is 50\% in case (a) and 40\%
in case (b).

The minimum masses given above depend on the
model used for the calculations, in particular the poorly-determined
velocity dispersions in the bulge population (Ibata \& Gilmore, 1995).
This yields the largest uncertainties in the present results,
since $\langle  m\rangle\propto \langle\sigma^2\rangle$.

These mass functions
yield a brown dwarf contribution to the surface density in the
solar neighborhood, assuming the same scale height as for the M-dwarf
population, i.e.  $h_d\sim 300$ pc:

\beq\Sigma_{\rm bd} \approx 3\pm 3 \ \mssurf\label{Sigbd1}\eeq

Therefore the brown dwarf contribution to the local
surface density is about 10\% to 20\% of the total stellar contribution
(\ref{Sigmastar2}). This represents $\sim 10$\% of the observed baryonic
mass (\ref{Sigobs}). As shown in Eq. (\ref{Sigbd1}), the brown dwarf
contribution is also compatible with zero. The {\it total} surface
density in the form of baryonic matter in the disk is thus
(Eqs. (\ref{Sigobs}) and (\ref{Sigbd1})):

\beq\Sigma_{\rm baryon} = 46\, \pm\, 6\,\mssurf\label{Sigbar}\eeq

\noindent consistent with the dynamical determination (\ref{SigFF}).
There is no need for any additional dark matter. The contribution of {\it
stellar and substellar objects} to the surface density is (Eq.(\ref{Sigmastar2}) and Eq.(\ref{Sigbd1})):

\beq\Sigma_{ss} = 30\, \pm \, 5\ \msol\, pc^{-2}\label{Sigss}\eeq

\noindent This yields an optical depth (see e.g. Kiraga \& Paczy\'nski, 1994):

\begin{eqnarray} \tau_{\rm disk} & \approx & {2\pi G\over 3 c^2}\rho_{*} R_\odot^2
\nonumber \\
 & \approx & 5.5 \times 10^{-7} \, \frac{\Sigma_{\rm ss}+50\mssurf}{50 \mssurf}
\nonumber \\
& \approx & 9.0 \pm  0.5 \times 10^{-7},\label{TauDisk}
\end{eqnarray}

\noindent i.e. $ \tau_{\rm disk} \wig < 10^{-6}$,
a factor $\sim$ 2 to 3 smaller than the observed value $\tau_{\rm
obs}\sim 2.4\,\pm\,0.5\,\times 10^{-6}$ (Alcock et al.,
1997).\nocite{Alcocketal97} Therefore, although the observed
time-distribution of the events is well reproduced by the present model,
the optical depth amounts to $\sim 1/3$ of the observed value. This
clearly suggests a bulge strongly elongated along the line-of-sight or,
similarly, a bar model (Zhao 1994).\nocite{Zhao94} We have not included
a bar-model in our calculations, but from the present analysis, its
contribution to the optical depth corresponds to:

\beq\tau_{\rm bar} \approx 1\times 10^{-6}\, M_{\rm bar}/(10^{10}
\msol)\label{Taubar}\eeq

\noindent in good agreement with various models (Zhao, Spergel
\& Rich, 1995, 1996; Han \& Gould,
1995; Stanek et al., 1997; Bissantz et al., 1997).
Infrared observations from DIRBE seem to constrain the mass of the
bulge within 2.4 kpc to $M_b\sim 7.2-8.6 \times 10^9\,\msol$, although
this value is obtained by substracting a model-disk contribution;
a more general constraint concerns the {\it bulge}+{\it disk} mass
$M_{b+d}(<2.4 \ \mbox{kpc}) =1.9\times 10^{10}\,\msol$ (Bissantz et al.,
1996).\nocite{Bissantzetal96} Note however that the three long-time
events contribute about 1/3 to the observed optical depth, so that the
discrepancy between the value (\ref{TauDisk}) and the observed one is
reduced appreciably if these events are removed, leading to a smaller mass
for the bulge. Moreover, as noted recently by Han (1997)\nocite{Han97}
and Alard (1997),\nocite{Alard97} the optical depth toward the bulge might
be overestimated by as much as a factor 1.7, because of amplification
bias due to pixel lensing.

\medskip

\subsubsection{Events whose source is a bulge clump giant}
\label{sourcegeante}

As mentioned by Alcock et al. (1997),\nocite{Alcocketal97} there
are significant uncertainties in the estimated bulge optical depth,
because essentially of blending effects. For this reason, the MACHO
collaboration has conducted a separate analysis of the events whose
source is a bulge {\it giant} (Alcock et al. 1997).\nocite{Alcocketal97}
Although the statistics are significantly diminished ($1.3\times 10^6$
source giants and 13 events), the efficiency and the reliability of the
results are improved appreciably. The corresponding optical depth is
also substantially larger, but only at the $\sim 1\sigma$ level, than
the value obtained when including the totality of the events, namely
$\tau_{\rm giant}\sim 3.9\times 10^{-6}$. The mean duration of these
events is $\langle  t_e\rangle=33.8$ days, which yields an unrealistic
average mass of 0.6 $\msol$ from relation (\ref{sqrtm}). This high
value is confirmed by the complete calculation, with a minimum mass
$\minf\approx 0.3\msol$. This minimum mass is the same for both MFs (a) and (b) because they are identical for $m>0.35\msol$. Such a high minimal mass is not allowed in the disk, since many stars with mass $m<0.3\ \msol$ are observed in the solar neighborhood. Therefore, either the bulge mass function is different from the disk MF, or the three long time scale events have a different
origin (most likely stellar remnants).

Without including the three afore-mentioned long-time events, which all
correspond to a bulge clump giant source, the mean duration is $t_e=16.7$
days, yielding an average mass of 0.15 $\msol$. The minimum mass derived
with the complete calculations to reproduce the time distribution in
that case is $\minf= 0.11\,\msol$ in case (a) and $\minf= 0.07\,\msol$
in case (b). The KS test gives a result of 70\% in both cases. Figure
\ref{figmod} shows the model distribution compared to observations in
case (a). It is thus possible to explain the observed time distribution
with a main sequence stellar population with the mass function (a) or (b)
plus a stellar remnant population.

\subsubsection{Events whose source is a disk main sequence star}

The OGLE collaboration (Paczi\'nski et al., 1994)\nocite{Paczynskietal94}
and the MACHO
collaboration (Alcock et al., 1997),\nocite{Alcocketal97} have
observed the same region, near the Baade's window, in the direction of the bulge. The MACHO collaboration observed {\it six}
events whose source lies on the {\it disk} MS, with a magnitude $V\wig <
18.5$. The sources of 3 of these 6 events lie near the very red edge
of the MS and could well be bulge blue helium core burning giants.
This could be easily verified from their spectra. However, the
remaining 3 sources are likely to be disk MS stars. These events yield
an optical depth $7.5\times 10^{-7}$, whereas the standard disk model
(with no dark population)
yields $\sim 10^{-7}$ and a {\it maximal disk} (see paper II) yields
$\sim 2\times 10^{-7}$. But the MACHO efficiency used for the derivation
of this optical depth is an average over all source stars. Since we
are considering only sources with $V<18.5$, the efficiency for these
brighter sources must be higher. With an extreme (unlikely) 100\% efficiency, the
optical depth for disk sources is $\tau=2\times 10^{-7}$ in better
agreement with disk models. We stress that theoretical estimates
depend strongly on the disk scale length and scale height.

The average time for the corresponding 3 events is $\langle
t_e\rangle=4.4$ days. Since the source stars are as bright as the
bulge clump giants, the blending is expected to be negligible for these
events. The mean duration would correspond to a minimum mass smaller
than 0.005 $\msol$, definitely in the brown dwarf domain. However,
the extrapolation of either mass function (a) or (b) down to a minimum
mass of 0.005 $\msol$ requires a moderate power-law exponent in order
not to exceed the local dynamical limit ($\Sigma_{\rm bd}=27 \,\mssurf$
if $\alpha=2$ and $\Sigma_{\rm bd}=62\,\mssurf$ if $\alpha=2.5$ for MF
(a)). Moreover, such a low minimum mass overproduces short time scale
events for bulge sources, especially when only clump giant sources
are considered.

\begin{figure} \resizebox{\hsize}{!}{\includegraphics{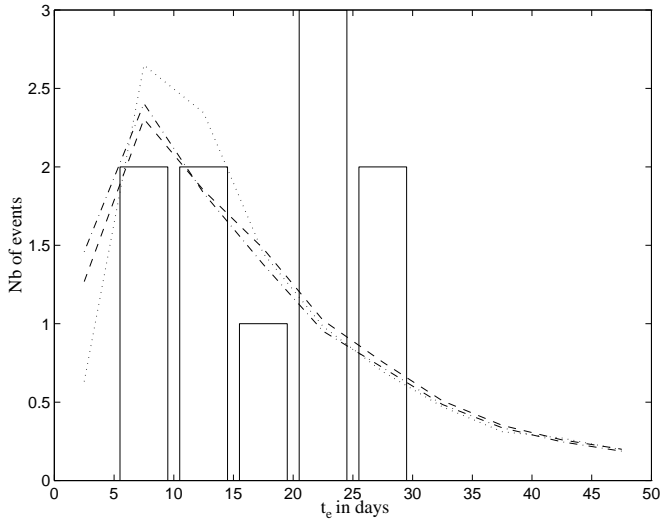}}
\caption{Comparison of the giant bulge events (excluding the 3 longest events)
with three models. Dotted line: the MF (a) is assumed to be the same
for both the bulge and the disk, with $\minf=0.11\,\msol$. Dashed
line: the disk MF has a $\minf=0.015\,\msol$ and the bulge MF has a
$\minf=0.4\,\msol$, for MF (a); the dot-dashed line is the same for MF
(b). The histogram corresponds to observations.\label{figmod}}
\end{figure}

Figure \ref{figmod} shows the calculated time distribution of the
microlensing events obtained with the MF (a) with {\it two different
minimum masses} for the bulge and the disk, as inferred from \S 4.1.1
and 4.1.2. A minimum mass $\minf=0.3\,\msol$ for the {\it bulge} and
$\minf=0.015\,\msol$ for the {\it disk}, with both MFs (a) and (b). The
model distributions agree with a 70\% KS result with the observed one.
Such a disk brown dwarf population corresponds to a local surface density:

\beq\Sigma_{\rm bd} \approx 20-30\, \mssurf\label{Sigbd2}\eeq

Clearly any solution between the above one and the one where $\minf\sim
0.1\,\msol$ everywhere (cf \S \ref{sourcegeante}), i.e. no brown dwarf, looks acceptable. The
disk contribution to the bulge optical depth in this model is $\tau_d
\approx 8\times 10^{-7}$, slightly larger than the optical depth of a
standard (no brown dwarf) disk ($5\times 10^{-7}$), but still insufficient
to explain the observed $\tau \approx 4\times 10^{-6}$.

\bigskip

Whereas small number statistics and uncertainty in the distance of the
sources prevent a reliable determination of the brown dwarf contribution
to the disk mass budget in this case, the present analysis corresponds to
a brown dwarf and a dynamical density in the disk, from (\ref{Sigobs}),
(\ref{Sigbd1}) and (\ref{Sigbd2}):

\beq \Sigma_{bd}\approx 0-30\, \mssurf\,\,\,\,\,\,\,\, ; \,\,\,\,\,\,\,
\Sigma_{\odot}\approx 40-70\,\mssurf,\label{Total}\eeq

\noindent possibly larger than
the value (\ref{SigFF}) determined in \S 2.1.
A mass-dependent scale height does not change
significantly the brown dwarf contribution, unless their
scale height is larger than $\sim 1$ kpc.

\medskip

These results leave open the following possibility: {\bf i)} the mass
functions of the disk and the bulge are different, in particular
in term of minimum mass, and {\bf ii)} the dynamical upper limit
for $\Sigma_\odot$ (Eq. \ref{SigFF}) and the observed density of
baryonic matter (Eq. \ref{Sigobs}) imply that 20 to 30 $\mssurf$
of the solar surface density can be in the form of brown dwarfs, in
agreement with the microlensing analysis (Eq. \ref{Total}).  Note that
the possibility for the disk and bulge mass functions to be different
has also been suggested by Gould et al. (1997) to reconcile star counts
and microlensing observations. However, although these authors invoke a
large population of brown dwarfs in the {\it bulge}, the present analysis
suggest that they could as well be located in the {\it disk}. A more
precise determination of the amount of brown dwarfs in the disk requires
microlensing observations optimized for characteristic timescales of a
few days, corresponding to a sampling of two observations per day. We
stress the need for such microlensing searches.

\medskip

The present analysis illustrates the difficulty to reach robust
conclusions about the mass under the form of sub-stellar objects in
the central regions of the Galaxy, and more precisely in the disk
and the bulge, from present microlensing experiments. The situation
will certainly be clarified once future projects will be operational,
allowing much better statistics in both parts of the Galaxy. In fact,
the MACHO survey has now a total of more than 200 candidate microlensing
events, still under analysis, which will improve
the present determinations.

The volume density of {\it disk} brown dwarfs in the solar neighborhood
deduced from the two types of calculations described above, global vs
separate disk/bulge analysis, correspond respectively to $\rho_{\rm
BD}\sim 5\times 10^{-3}\,\msvol$ (from Eq. \ref{Sigbd1}) within
about a factor 2, and $\rho_{\rm BD}\sim 3-5 \times 10^{-2} \,\msvol$
(from 3 events in the disk, Eq. (\ref{Sigbd2}): this is clearly to be
revised in the light of future observations), a factor 10 more. Future
deep photometry and large field observational surveys might thus help
determining the present issue by direct observation of nearby brown
dwarfs, although the small density and the intrinsic faintness of brown
dwarfs ($M_V\wig > 22$, Baraffe et al. 1997)\nocite{Baraffeetal97}
renders the observation of a statistically-significant sample of field
brown dwarfs a tremendously difficult task. Note that infrared filters
are highly recommended for these objects ($M_J\wig > 12$; $M_K\wig >
11$, Baraffe et al. 1997).\nocite{Baraffeetal97}

The present analysis is thus an indication that an important amount of
brown dwarfs in the disk is not excluded with present day observations, as
suggested by the rise of the disk mass function near $0.1\msol$ 
(cf. \S 3.1).

\subsection{Spheroid}

For a DeVaucouleurs spheroid, extrapolation of the MF (\ref{SpheMF})
into the brown dwarf region, even with $\minf=0.01\,\msol$, yields
a number of microlensing events towards the LMC of $\sim$ 0.4 and an
optical depth $\tau_{\rm sph} \sim 5\times 10^{-9}$ (Chabrier \& M\'era,
1997).\nocite{ChabrierMera97} This illustrates the negligible contribution
of the spheroid to the Galactic mass, and thus to the events observed
towards the LMC.

\subsection{Dark Halo}

The analysis of the {\it first year} of the EROS (Aubourg
et al., 1993)\nocite{Aubourgetal93} and MACHO (Alcock et al.,
1993)\nocite{Alcocketal93} microlensing observations towards the LMC had
shown that the observed events were likely to be due to halo brown dwarfs,
with an average mass $\langle  m\rangle\approx 0.03\,\msol$ (M\'era et
al. 1996b; Kerins 1995).\nocite{MeraChabrierSchaeffer96,Kerins95}  The
inferred contribution of these objects to the missing mass was found to be
between 10 and 20\% (Alcock et al. 1995; Gates et al., 1995; M\'era et al.
1996b).\nocite{Alcocketal95b,GatesGyukTurner95c,MeraChabrierSchaeffer96}
These calculations must now be re-examined in the context of the {\it
second year} of the MACHO experiment.

These results yield now a total of 1 to 2 events observed by EROS (Ansari
et al., 1996; Renault et al., 1997)\nocite{Ansarietal96,Renaultetal97}
and 6 to 8 by MACHO (Alcock et al., 1996).\nocite{Alcocketal96lmc}
The inferred optical depth is much larger than derived previously,
$\tau_{\rm obs} = 2.2\,\pm 1\times 10^{-7}$, about 40\% of the value
corresponding to the standard halo.  The 6 MACHO events\footnote{We
have excluded the MACHO events \#10, believed to be a variable star
(Alcock et al., 1996),\nocite{Alcocketal96lmc} and \#9, the binary event,
which probably belongs to the LMC.} yield $\langle  t_e\rangle\approx 40$
days, in reasonable agreement with the value inferred from the 2 events
observed by EROS, $\langle  t_e\rangle\approx 28$ days. The statistics,
however, are not sufficient to constrain a mass function from the
time distribution. As will be shown in Paper II, the number of observed
events cannot be explained by dark objects in the LMC itself. As shown
by Chabrier \& M\'era (1997)\nocite{ChabrierMera97} and Graff \& Freese
(1996),\nocite{GraffFreese96} extrapolation of the MF (\ref{SpheMF})
into the brown dwarf domain yields a negligible fraction of brown dwarfs
in the Galactic halo.

A substantial amount of these objects in the halo thus implies a
significantly different, much steeper MF beyond the hydrogen-burning
limit. In order to estimate the {\it maximum contribution} of halo
substellar objects to the microlensing counts, we consider, as in M\'era
et al (1996b),\nocite{MeraChabrierSchaeffer96} a halo MF $\frac{dN}{dm}(m)
\propto m^{-x}$ with $x = 5$, essentially a Dirac-peaked MF. The {\it
maximum normalization} of $\frac{dN}{dm}(m)$ at 0.1 $\msol$ is the
observed count of halo subdwarfs in the solar neighborhood, derived in \S
3.4. We thus take as the {\it upper limit} for the dark halo substellar
mass function:

\beq\frac{dN}{dm}(m) = 4.0\times 10^{-3} \left(\frac{m}{0.1
\msol}\right)^{-5} \msol^{-1} \mathrm{pc}^{-3}\label{HaloMF}\eeq

A Kolmogorov-Smirnov test done on the $\te$ distribution
of the EROS + MACHO halo events shows unambigously that such a MF is
not consistent with the observations, whatever the value of
$\minf$. The optical depth can be reproduced only if $\minf<0.02\msol$,
but the time distribution is in severe conflict with the observed one,
as shown by a 0.02\% KS probability ($\langle  \te\rangle=12$ days).

The expected average mass of microlensing events is related to
the characteristic timescale of the event and to the velocity
dispersion (see Appendix Eqs. A2 and A15). The average tangential
velocity is proportional to the dispersion of the Gaussian velocity
distribution $\langle  1/v_\perp\rangle^{-1}\propto \sigma$ (see
Appendix). Assuming a {\it maximal} dark halo $\rho_h(r)=\rho_{\rm
dyn}/r^2$, where $\rho_{\rm dyn}=8\times 10^{-3}\,\msvol$ is the
local dark matter density (Bahcall 1986),\nocite{Bahcall86} this
dispersion corresponds to the asymptotic rotation velocity with
the relation $\sigma= v_{\rm rot}/\sqrt 2\approx 150\,\kms$, where
$v_{\rm rot}\approx 220$ km.s$^{-1}$, as given by the isothermal
sphere model (Binney and Tremaine 1987).\nocite{BinneyTremaine87}
This yields an average mass, corrected for blending, $\langle  \sqrt
m\rangle^2\approx 0.4-0.5\,\msol$, as suggested by the MACHO collaboration
(Alcock et al., 1996).\nocite{Alcocketal96lmc} Several observations
(Beers \& Sommer-Larsen, 1995; Dahn et
al., 1995; Layden et al., 1996) suggest that
the halo population velocity ellipsoid is radially elongated, by a
factor $\sim 1.5-2$. This yields a projection effect which decreases the
velocity dispersion. On the other hand, this dispersion is bound by the
velocity of the line-of-sight toward the LMC (Eq. \ref{vrel}), which is
(after proper calculations), $\langle  1/v_{\rm los}\rangle^{-1} \sim
80\,\kms$. This corresponds to a minimum mass $\minf = 0.04\,\msol$ but
to a {\it zero} dispersion velocity, clearly an irrealistic possibility.
When considering such a ``modified'' halo model, with a non-isotropic
velocity dispersion tensor at 1$\sigma$ of 80 $\kms$ and an important
oblateness, the time distribution can be reproduced with $\minf\sim
0.1\msol$, still above the hydrogen-burning limit. However, the number of
observed events implies a normalization for the halo MF at 0.1 $\msol$
significantly larger than the maximum value (\ref{SpheMF}), inferred
from observed low-mass stars in the solar neighborhood.

These calculations show that, if the results of the recent MACHO
analysis are confirmed, the possibility for halo dark objects to be field
brown dwarfs is clearly excluded, even for a MF substantially steeper
than the ones in the disk and the spheroid, for any halo model. This
analysis, consistent with the MACHO observations (Alcock et al.,
1996)\nocite{Alcocketal96lmc} and with star count analysis at faint
magnitudes (Chabrier \& M\'era, 1997),\nocite{ChabrierMera97} and the
one conducted in \S3.4 and 3.5, suggest a negligible mass-fraction
under the form of field stars and brown dwarfs in the dark halo.
Different solutions to try to reconcile star counts and microlensing
observations in the halo will be considered in paper II.

\section{Summary and conclusions}

In this paper, we have examined in detail all the observational
sources that constrain the mass distribution of the Galaxy. This
includes dynamical constraints, circular rotation velocity and vertical
velocity dispersion in the solar neighborhood, star counts in the disk,
bulge and spheroid, and microlensing observations towards the Galactic
halo and the Galactic center. The star counts yield the derivation of
stellar mass functions for the disk, the bulge and the spheroid down to
the bottom of the main sequence.  A substantial discrepancy, however,
remains between the MF inferred from the nearby LF and the one deduced
from the HST LF in the range $0.09-0.35\,\msol$. Although the nearby
MF is consistent with a steadilly rising power-law $dN/dm\propto
m^{-\alpha}$ with $\alpha \sim 2$ down to the bottom of the main sequence,
the HST MF is essentially flat below 0.6 $\msol$.
Interestingly enough, the significantly rising last bin of the HST MF,
indicates an upturn of the MF near the brown dwarf domain, so that both
type of MFs, nearby and photometric, suggest the presence of a substantial
amount of brown dwarfs in the Galactic disk. 

We have determined the normalization of these mass functions and shown
that the ratio of the halo to disk main sequence stellar density is
about $\sim 1/400$ near the bottom of the main sequence. The presence
of the spheroid is proven unambigously by the observed star counts
of high velocity stars but its contribution to the Galactic mass is
negligible. These calculations yield the determination of the contribution
of main sequence stars, and thus of {\it observed} baryonic matter, to
the disk and halo mass budget. The detected (thin+thick) disk stellar
surface density corresponds to $\Sigma_{vis}\approx 43\,\pm\,5\,\mssurf$,
whereas the halo (spheroid + dark halo) density corresponds to less than
$\sim 1\%$ of the dynamically determined dark matter, significantly less
than the limit determined by the HST (Flynn et al., 1996).

The contribution of sub-stellar objects to this mass depends on the lower
limit down to which the mass functions can be extrapolated. This limit
is fixed by the microlensing observations, in particular by the time
distribution of the events. This combined analysis of star counts for the
luminous part of the mass function, which yields the determination of the
slope and normalization near the brown dwarf limit, and of microlensing
experiments, which determines the minimum mass, yields a {\it consistent
determination} of the dark matter under the form of star-like objects,
low-mass stars and brown dwarfs, in different parts of the Galaxy.
Although uncertainties due to either small statistics or ill-constrained
velocity dispersion and location of the sources in the bulge prevent for
now a precise determination, the derived brown dwarf surface density in
the solar neighborhood, for a scale height $h_z=320$ pc (Gould et al.,
1997), is likely to be in the range $\Sigma_{\rm bd}=0-30 \,\msol.pc^{-2}$
so that the total local density in the disk is $\Sigma_{\rm dyn}\sim
40-70\,\mssurf$. As will be shown in Paper II (M\'era et al., 1997),
this is in agreement with various observational determinations. More
statistics of microlensing events towards the bulge will allow a separate
determination of the brown dwarf contribution in the bulge and in the
disk. An attempt based on the analysis of the events whose source is
a clump giant or a disk main sequence star, respectively, yield rather
surprising results (hampered by present observational uncertainties):
most brown dwarfs would be present in the disk, whereas the bulge would
contain almost none of these objects. This stresses the need for further
observations in this region.

For the halo, we show that brown dwarfs have a negligible contribution
($\wig < 1\%$) to the halo mass. A steep ($\alpha \wig > 2$) mass
function in the halo seems to be excluded both by the microlensing
analysis and by the M-dwarf LF. Therefore, the present analysis excludes
main sequence stars and brown dwarfs as a significant contribution to
the halo mass budget.

The nature of the dark events observed in the halo remains a puzzle.
Halo white dwarfs remain the least unlikely candidates, although
this scenario implies severe constraints on the age of the halo and
its initial mass function (Chabrier et al., 1996; Adams \& Laughlin,
1996).\nocite{ChabrierSegretainMera96,AdamsLaughlin96} More attention
will be devoted to this problem in Paper II.

\medskip

These calculations, which combine observational constraints arising
from star counts, microlensing experiments and kinematic properties,
yield the consistent determination of the amount of dark matter under
the form of stellar and sub-stellar objects in the differents parts of
the Galaxy. This yields new insight on the distribution of baryonic dark
matter in the Galaxy and bears important consequences for the derivation
of a consistent Galactic mass-model. This will be examined in the next
paper (M\'era, Chabrier and Schaeffer, 1997, paper II).\nocite{Meraetal97b}

\bibliographystyle{astron}

\appendix
\onecolumn

\section{Microlensing equations}

The basic microlensing equations can be found for instance in Griest
(1991) or in Kiraga and Paczy\'nski (1994). However, we use some extra
formulae which are given below, after a summary of the basic equations.

The duration $t$ of a microlensing event is defined as the time during
which the amplification of the monitored star is larger than a given
threshold amplification $A_T$\footnote{or equivalentelly, a threshold impact
parameter $u_T$ in units of the Einstein radius, related to $A_T$
by $\displaystyle A_T = \frac{u_T^2+2}{u_T\sqrt{u_T^2+4}}$.} (usually
$A_T=1.34$). The event duration corresponds to the crossing time of the
Einstein disk, of radius $u_T R_e$, for the lens:

\beq t=\frac{2R_e}{v_\bot}\sqrt{u_T^2-u_{min}^2}=53\
\mathrm{days}\ \frac{220 \mathrm{km.s}^{-1}}{v_\bot}\times
\sqrt{\frac{m}{0.1\msol}} \sqrt{\frac{L}{55\, \mathrm{kpc}}}
\frac{\sqrt{x(1-x)}}{0.5}\sqrt{u_T^2-u_{min}^2}\label{t}\eeq

\noindent where $v_\bot$ is the lens transverse velocity w.r.t. the line
of sight, $u_T$ (resp. $u_{min}$) is the impact parameter
corresponding to the threshold amplification (resp. maximum
amplification) $A_T$, $L$ is the distance to the source, $xL$ and $m$
denote respectively the distance and the mass of the lens, and
$R_E=\frac{2}{c}\sqrt{GmLx(1-x)}$ is the Einstein radius.

The {\it characteristic} time of an event is defined as:

\beq t_e=\frac{R_e}{v_\bot} = \frac{2}{cv_\bot}\sqrt{GLmx(1-x)} =
\frac{t}{2\sqrt{u_T^2-u_{min}^2}}\label{te}\eeq

This effective time does not depend on the impact parameter, and can
be recovered from the observations with the relation between $u_{min}$
and the known maximal amplification. In practice, the blending (several
unresolved stars of which only one is amplified) renders this process rather difficult. The overall effect of blending is to
underestimate the characteristic time.

The probability for a source star to be microlensed at a given time is called the optical depth, and reads:

\beq \tau= \int_0^1 u_T^2 \pi R_e^2 \frac{\rho(xL)}{m} L\dx
= \int_0^1 u_T^2 \pi \frac{4GL^2}{c^2} \rho(xL) x(1-x) \dx
\label{tau}\eeq

\noindent where $\rho(xL)=\int_0^\infty \rho_m(xL) dm$ is the total
mass-density under the form of dark objects.

The experimental optical depth is retrieved from observations by:

\beq\tau_{exp}=\frac{\sum t_{obs}/\epsilon(t)}{N_s\times
T_{obs}}\label{tauexp}\eeq

\noindent where $N_s$ and $T_{obs}$ denote the number of source stars
and the total duration of the experiment, respectively whereas $\epsilon(t)$ is
the detection efficiency.
\bigskip

\indent$\bullet$ {\bf event rate:}

The theoretical event rate for a given Galactic model, i.e. the expected
number of events per unit time, reads:

\beq\rmd\Gamma = 2u_TR_ev_\bot \frac{\rho(xL)}{m}
\rmP(m)\rmP(v_\bot)\dm\dv_\bot\dx\label{dGamma}\eeq where P($m$) and
P($v_\perp$) are the probability distributions respectivelly of lens
mass and velocity.  If the velocity distribution is independent of the
position, the integration of (\ref{dGamma}) yields:

\beq\Gamma  =  u_T\frac{4}{c}\sqrt{GL}
\int_{\minf}^{m_{sup}}\frac{1}{\sqrt{m}}\rmP(m) \dm
\times  \int_0^L\sqrt{x(1-x)}\rho(xL) \rmd (xL) \int_0^\inf
\dv_\bot v_\bot \rmP(v_\bot)\label{Gamma}\eeq

For $N_s$ source stars monitored during $T_{obs}$, the expected {\it
number} of events is:

\beq N=\Gamma \times N_s T_{obs} \label{expnum}\eeq

The relation (\ref{dGamma}) shows that the {\it effective} microlensing
probability distributions are different from the model distributions,
and are given by:

\beq\rmP_{eff}(x)   \propto  \sqrt{x(1-x)} \rho(xL)\label{Peffx}\eeq

\beq\rmP_{eff}(v_\bot)  \propto  v_\bot \rmP(v_\bot)\label{Peffv}\eeq

\beq\rmP_{eff}(m)  \propto  \frac{\rmP(m)}{\sqrt{m}}\label{Peffm}\eeq

For a Maxwellian velocity distribution, of dispersion $\sigma$, the
transverse velocity reads:

\beq\rmP(v)\dv = \frac{1}{2\pi
\sigma^2}e^{-\frac{v_x^2+v_y^2}{2\sigma^2}}\dv_x\dv_y =
\frac{v}{\sigma^2} e^{-\frac{v^2}{2\sigma^2}}\dv\label{Pvgaus}\eeq

\noindent and

\beq\rmP_{eff}(v)=\frac{v\rmP(v)}{\int_0^\inf v'\rmP(v') \dv'} =
\sqrt{\frac{2}{\pi}}\frac{v^2}{\sigma^3}
e^{-\frac{v^2}{2\sigma^2}}\label{Peffvgaus}\eeq

\noindent which yields:

\beq\rmP(t_e) = \frac{ \int \frac{\rmP(m)}{\sqrt{m}}
\sqrt{x(1-x)}\rho(xL) \sqrt{\frac{2}{\pi}} \frac{R_e^2}{t_e^2 \sigma^3}
e^{-\frac{R_e^2}{2\sigma^2 t_e^2}} \frac{R_e}{t_e^2} \dx\dm}{ \int
\frac{\rmP(m)}{\sqrt{m}} \sqrt{x(1-x)}\rho(xL)\dm\dx}\label{Pte1}\eeq

\noindent and (with use of eqn. (\ref{Gamma})):

\beq\rmP(t_e)=\frac{32G^2L^3u_T}{c^4t_e^4\Gamma\sigma^2}\times
\int m\rmP(m) \left[ x(1-x) \right]^2 \rho(xL)
e^{-\frac{2GLmx(1-x)}{c^2\sigma^2t_e^2}}\dx\dm\label{Pte2}\eeq

For a given Galactic model, we use a Monte-Carlo algorithm to calculate the event rate, which yields a set of simulated events whose time distribution is exactly the one given by (\ref{Pte2}).

These equations yield the average characteristic time $\langle  t_e\rangle$:

\beq\langle  t_e\rangle=\frac{2\sqrt{GL}}{c}\langle  \sqrt{m}\rangle\langle  \frac{1}{v_\bot}\rangle\langle  \sqrt{x(1-x)
}\rangle\label{temoy}\eeq

In equation (\ref{temoy}), the means are computed with the effective
probability distributions (\ref{Peffx}-\ref{Peffm}). When writing
explicitely the corresponding integrals, we can identify the optical
depth and the event rate, which yields:

\beq\tau = \Gamma\times\frac{\pi}{2}u_T\langle  t_e\rangle.\label{tau_te_rel}\eeq

A given Galactic model implies a time-distribution
$d\Gamma/dt_e=\Gamma\times P(t_e)$ and the number of events predicted
by the theory is given by:

\beq N_{th}=E\times \int_0^{+\infty} \epsilon (t_e)
\frac{\mathrm{d}\Gamma}{\mathrm{d}t_e} \mathrm{d}t_e \label{NthA}\eeq

where $E=N_s\times T_{obs}$.
\bigskip

\indent$\bullet$ {\bf Divergence of the moments:}

As shown in Eq. \ref{Pte1}, the probability to observe an effective time $t_e$ behaves as :

\beq P(t_e) dt_e \propto 1/t_e \, P(v_\bot) dv_\bot \propto v_\bot^3 dv_\bot \eeq

so that $P(\te)$ has a long power-law tail at large times.
This yields the divergence of the moments of order $n>3$ $\int t_e^n P(t_e) dt_e$, which causes the extreme sensitivity of $\langle  t_e\rangle$ and $\tau$ to the rare large-time events. This long-range tail effect, which prevents the use of the higher order moments of $\te$, is also felt when the central limit theorem is
invoked to identify the {\it observed} average time of e.g. the 45 events
with the {\it statistical} average. They are equal in the large-N limit, but the convergence is extremely slow in this case. This shortcoming is fixed by using the average
$\langle  {1\over \te}\rangle$ instead of $\langle  \te\rangle$. This method is more useful to constrain $\minf$ and $\langle  \te\rangle$ than
Eq .\ref{temoy} since $1/\te\propto v_\bot$ and is
thus nearly gaussian at the beginning and  $1/\te\propto m^{-1/2}$ and is thus most sensitive to the low mass end of the mass function.

\bigskip

$\bullet$ {\bf Motion of the line of sight:}

The lens velocity w.r.t. the line-of-sight (l.o.s.) depends on the motion
of the l.o.s. itself and must include the motion of the Sun and of the
source star w.r.t. the Galactic center. The lens velocity thus reads:

\beq\vec{v}=\vec{v}_{lent.}-x\vec{v}_{source}-(1-x)\vec{v}_\odot
\label{vrel}\eeq

The velocity now depends on the lens distance $xL$, and equations
(\ref{Pvgaus}-\ref{Pte2}) relative to a gaussian velocity distribution,
are no longer valid. Straightforward calculations yield the projected
velocity of the lens w.r.t. the l.o.s.:

\beq v  =  \left[ \left( v_{lent.,l}-v_{\odot ,l} +x(v_{\odot ,
l}-v_{s,l})\right)^2 + \left( v_{lent.,b}-v_{\odot ,b} +x(v_{\odot ,
b}-v_{s,b})\right)^2 \right]^{1/2}\label{vrellb}\eeq

\bigskip

$\bullet$ {\bf Distance of the source star:}

In the case of the bulge, the elongation along the line of sight is no
longer negligible, and the afore-mentioned formulae must include the
possible variation of the distance $L$ of the source, which implies an
extra integral on $L$:

\beq\tau=\frac{ \int_0^\inf L^2 \nu_s(L) \d L \int_0^1 u_T^2 \pi
\frac{4GL^2}{c^2} \rho(xL) \sqrt{x(1-x)} \dx}{\int_0^\inf L^2 \nu_s(L)
\d L}\label{tau_dvar}\eeq

\noindent where $\nu_s$ is the density of source stars {\it visible}
at the distance $L$.  This requires a luminosity function for the
source stars. Following Kiraga and Paczy\'nski (1994), $\nu_s \propto
\rho L^{2\beta}$, if the number of stars brighter than some absolute
luminosity L is proportional to L$^\beta$. We have slightly modified
$\nu_s$ to take into account the fact that giant stars have more or less
the same luminosity. Then $\nu_s \propto \rho(L^{2\beta}+C)$ where $C$
is adjusted to reproduce the observed ratio of giants.

The event rate (\ref{Gamma}) now reads:

\beq \Gamma = \frac{4\sqrt{G}}{c\int_0^\inf \nu_s(L) L^2 \d L}
\int_0^\inf \rmd m \int_0^L \rmd L \int_0^1 \dx \int_{-\infty}^\inf
\rmd^2 v_{lent} \int_{-\infty}^\inf \rmd^2 v_{S} \nonumber \\ \times
\frac{\rmP(m)}{\sqrt m} \ v_\bot\ \nu_s(L) L^{3.5} \rho(xL) \sqrt{x(1-x)}
\label{Gamma_dvar}\eeq This integral has six non-independent variables,
only the mass can be separated. The Monte-Carlo integration method (see
Press et al. 1992) does not rely on any discretization of the integration
domain, and is hence suitable for this kind of integral. Moreover, the
Monte-Carlo process provides a set of simulated microlensing events, from
which the time distribution can be recovered easily. The adjunction
of the experimental efficiency is also straightforward with a rejection
algorithm.

\end{document}